# An Accurate Interconnect Test Structure for Parasitic Validation in On-Chip Machine Learning Accelerators

CHUN-CHEN LIU, OSCAR LAW AND FEI LI

# Abstract

For nanotechnology nodes, the feature size is shrunk rapidly, the wire becomes narrow and thin, it leads to high RC parasitic, especially for resistance. The overall system performance are dominated by interconnect rather than device. As such, it is imperative to accurately measure and model interconnect parasitic in order to predict interconnect performance on silicon. Despite many test structures developed in the past to characterize device models and layout effects, only few of them are available for interconnects. Nevertheless, they are either not suitable for real chip implementation or too complicated to be embedded. A compact yet comprehensive test structure to capture all interconnectparasitic in a real chip is needed. To address this problem, this paper describes a set of test structures that can be used to study the timing performance (i.e. propagation delay and crosstalk) of various interconnect configurations. Moreover, an empirical model is developed to estimate the actual RC parasitic. Compared with the state-of-the-art interconnect test structures, the new structure is compact in size and can be easily embedded on die as a parasitic variation monitor. We have validated the proposed structure on a test chip in TSMC 28nm HPM process. Recently, the test structure is further modified to identify the serious interconnect process issues for critical path design using TSMC 7nm FF process.

# Table of Contents

# Chapter 1. Introduction

With the advent of big-data era, machine learning has become increasingly powerful in solving problems from various domains such as face recognition in security screening and high-frequency trade in banking. However, most machine learning algorithms are complex in nature and limited to real-time operation with software implementation. Accordingly, the hardware accelerated or learning on-chip [1-3] are evolved to enable deep learning applications [4-5], making use of powerful heterogeneous hardware platforms involving graphics processing units (GPUs), field-programmable gate arrays (FPGAs) and/or network-on-chips (NoCs). For example, Intel last year released Xeon processors with built-in FPGAs dedicated for data center and learning applications [6].

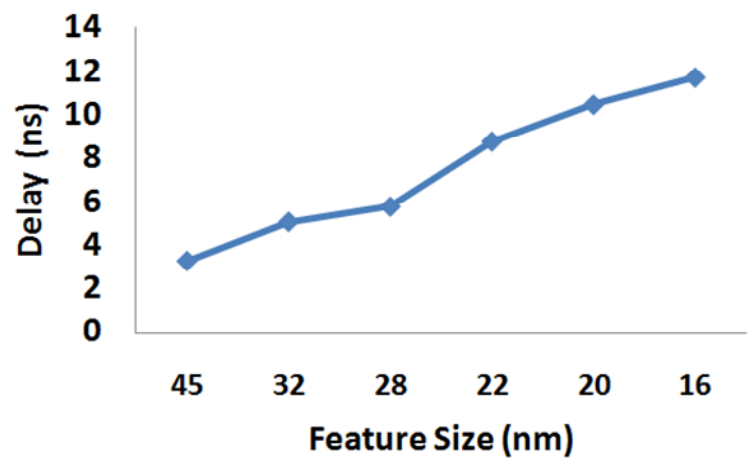

*Figure 1 Delay of 1-mm global interconnect in different technology nodes based on data reported in International Technology Roadmap for Semiconductors (ITRS) [7]*

With the evolution of CMOS technology, the performance of CMOS circuits (including analog [8-11], mixed-signal [12-15], RF [16-19] and digital [1-5]) has been improved profoundly. This is mainly due to the scaling of semiconductor device. However, the routing in CMOS circuits can hardly scale [7].A major bottleneck in these heterogeneous platforms lies in interconnects between various system components, as will be demonstrated in Section 2, the data latency is the decisive factor of the overall performance. On the other hand, with the relentless technology scaling, the features size is shrunk rapidly, the wire width and spacing are all significantly reduced, resulting in high coupling capacitance. To minimize the crosstalk impact, the wire thickness is also reduced with the major drawback of high sheet resistance due to a smaller cross-section area. The high sheet resistance leads to serious issue for Power Distribution Network (PDN) and clock tree design. From Figure 1, it can be seen that the interconnect delay increases drastically with technology scaling, the delay is increased several times from CMOS 45nm to 16nm process, it becomes more serious for 10nm and below.

Accordingly, it is crucial to develop an accurate interconnect model for accurate system performance simulation and prediction. Currently, many test structures [20-24] have been developed to characterize the impacts of standard cell architecture, custom datapath, current mirror and I/O pad design. However, only a few test structures exist with a focus on the impact of interconnect. In addition, in order to be useful and practical, they must be calibrated with measured data, to establish silicon-to-model correlation [25]. The conventional cross-bridge Kelvin structure [26-27] needs extra probe pads to directly measure the on-chip parasitic effect, which is not suitable for interconnect monitoring in real chip designs. Then, simple ring oscillator [28-29] is developed to measure the frequency of various interconnect configurations. This approach is widely adopted by the foundries and fabless design houses to validate the interconnect performance. However, it mainly focuses on the single interconnect and ignores the effect of cross coupling between adjacent wire impacts toward overall performance. Also, on-chip interconnect monitoring based on time-to-digital converter (TDC) has been proposed [30], but it suffers from the non-idealities of TDC. A compact yet comprehensive test structure to capture all interconnect parasitic is still a missing piece in the literature.

To address this issue, this paper describes a test structure based on a set of enhanced ring oscillator designs. It not only measures the propagation delay of various interconnect structures but also accounts for different crosstalk impacts (i.e. in-phase and out-of-phase crosstalk). Compared with current interconnect test structure, this proposal can measure both interconnect delay and crosstalk impact at the same time to significantly reduce the test structure area. It is easy to embed in the real chip for interconnect validation during the production. A first-order empirical model is also put forward to estimate the RC parasitic for silicon-to-model correlation. Since the test structure is relatively simple and small, it is easy to implement in real chips to monitor actual RC parasitic variations during manufacturing. We have validated the proposed structure on a TSMC 28nm test chip, and showed its efficacy on our in-house Neuro Processing Unit (NPU).

# Chapter 2. Interconnect Structures and Models

In this section, we briefly review the interconnect structures and models as they directly relate to our proposed test structures.

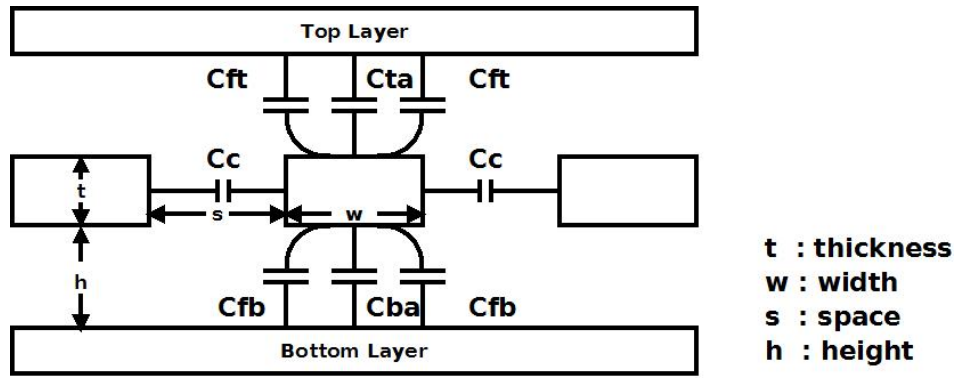

*Figure 2 Interconnect model*

The interconnect model where the wire is routed through top and bottom layer and coupled with two left/right adjacent wires is shown in Figure 2. The capacitances between the layers are named area capacitance ($C_a$) and fringe capacitance ($C_f$). They are further grouped into top ($C_{top}$) and bottom ($C_{bottom}$) capacitance as follows:

$$C_{top} = C_{ta} + 2\, C_{ft} \qquad \text{Eq. 1}$$
$$C_{bottom} = C_{ba} + 2\, C_{fb} \qquad \text{Eq. 2}$$

where $C_{ta}$ is top area capacitance, $C_{ba}$ is bottom area capacitance, $C_{ft}$ is top fringe capacitance, and $C_{fb}$ is bottom fringe capacitance.

The total capacitance is defined as the sum of the top, bottom and coupling capacitance ($C_c$) as

follows:

$$C_{total} = C_{top} + C_{bottom} + 2C_c \qquad \text{Eq. 3}$$

When wires run parallel to each other, the signal (victim) propagation is affected by the adjacent wires (aggressors) through the coupling capacitance as shown in Figure 3. It is called crosstalk impacts.

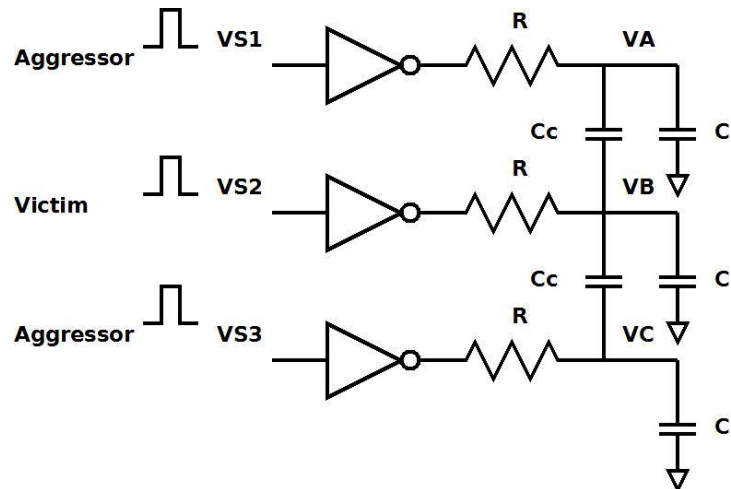

*Figure 3 Crosstalk model*

The crosstalk is highly dependent on the spacing between the adjacent wires (victim and aggressors) as well as the traveling directions. The effective coupling capacitance is significantly reduced when the spacing is increased. It also changes with different traveling directions due to the Miller effect. If the aggressor remains constant voltage, the victim is said to be quiet or shielded without any crosstalk impact. If the signals in both victim and aggressors travel in the same direction, the crosstalk is smaller (called in-phase crosstalk). If they travel in opposite directions, the effective coupling capacitance is larger (called out-of-phase crosstalk).

Assuming there is no switching in top and bottom routing layers, the left and right aggressor introduces the noise on the victim through the coupling capacitance ($C_c$). The effective

capacitance ($C_{eff}$) of the victim can be approximated using the following equations:

| | | |
|---|---|---|
| Ground capacitance | $C_{gnd} = C_{top} + C_{bottom}$ | Eq. 4 |
| Quiet mode | $C_{eff} = C_{gnd} + 2C_c$ | Eq. 5 |
| In-phase crosstalk | $C_{eff} = C_{gnd}$ | Eq. 6 |
| Out-of-phase crosstalk | $C_{eff} = C_{gnd} + 4C_c$ | Eq. 7 |

The crosstalk can be modeled using equivalent lump model in the Laplacian domain [30]. In order to simplify the model, all resistance $R$, capacitance $C$ and coupling $C_C$ are the same for all branches. $V_{S1}$ and $V_{S3}$ are the aggressor input voltage while $V_{S2}$ is the victim voltage. The output voltages $V_A$, $V_B$ and $V_C$ are defined as follows:

$$V_A = \frac{(1 + a_1 s + a_2 s^2)V_{s1} + (a_3 s + a_4 s^2)V_{s2} + a_5 s^2 V_{s3}}{(1 + b_1 s)(1 + b_2 s)(1 + b_3 s)} \quad \text{Eq.8}$$

$$V_B = \frac{a_6 V_{s1} + (1 + a_7 s)V_{s2} + a_8 V_{s3}}{(1 + b_4 s)(1 + b_5 s)} \quad \text{Eq.9}$$

$$V_C = \frac{a_5 s^2 V_{s1} + (a_3 s + a_4 s^2)V_{s2} + (1 + a_1 s + a_2 s^2)V_{s3}}{(1 + b_1 s)(1 + b_2 s)(1 + b_3 s)} \quad \text{Eq.10}$$

where

$a_1 = 2RC + 3RC_c$
$a_2 = R^2C^2 + R^2C_c^2 + 3R^2CC_c$
$a_3 = RC_c$
$a_4 = R^2C_c^2 + R^2CC_c$
$a_5 = R^2C_c^2$
$a_6 = RC_c$
$a_7 = RC + RC_c$
$a_8 = RC_c$

$b_1 = RC$
$b_2 = RC + RC_c$
$b_3 = RC + 3RC_c$
$b_4 = RC$
$b_5 = RC + 3RC_c$

In order to examine different crosstalk operations, $V_B$ is set to be a step function (with Laplacian transform of V/s) and $V_A$, $V_C$ are either set at zero for quiet mode operation or a step function (with Laplacian transform of ±V/s) with different polarity to model in-phase and out-of-phase crosstalk. It is then transformed to time domain and simplified through first order approximation.

# Chapter 3. Proposed Test Structure

## 3.1 Ring Oscillator Configuration

In this paper, we propose an enhanced test structure derived from ring oscillators. The overall structure is shown in Figure 4. It consists of an input control unit and three sets of ring oscillator. The control unit controls the test signal (i.e., victim) and two left/right adjacent routing signals (i.e., aggressors)for different crosstalk operations (i.e., quiet mode, in-phase and out-of-phase crosstalk). Finally, the test signal is fed into the down counter to scale down the ring oscillator frequency for oscilloscope measurement.

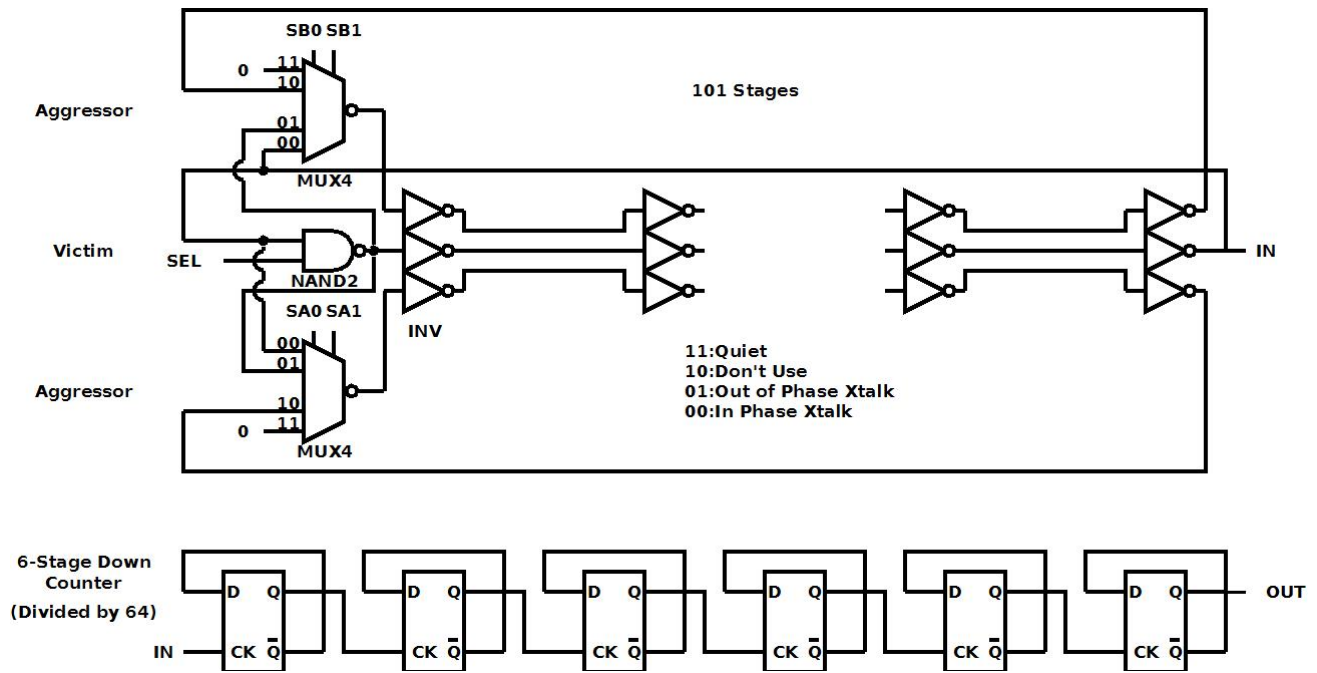

*Figure 4 Interconnect Test Structure*

The control unit is further divided into NAND2 and 4-inputs MUX; NAND2 is used to control the test signal (victim) propagation. When the control signal SEL is low, the output of NAND2 is always set to high which results in no oscillation.Whenthe control signal SEL is high, the output of NAND2 is inverted by the input signal IN and the signal propagates through the inverter chain and creates oscillation.

The aggressors are controlled by two 4-inputs MUX with in-phase crosstalk (00), out-of-phase crosstalk (01), quiet mode (11) and don't use (10)operations. If the select signals SAx and SBx are set to "10", it is defined as "don't use" or invalid state.If they are set to "11", it models the signal propagating alone through the inverter chain with shielded protection while the aggressors are set to one or zero to avoid any crosstalk impacts. If they are set to either "00" or "01", the input/output of NAND gates are fed into MUX to toggle the adjacent routing signals. The signals travel in the same or opposite directions as test one for in-phase and out-of-phase crosstalk study. The spacing between the victim and aggressors can be further adjusted to examine the various spacing crosstalk impacts toward signal propagation.

The proposed test structure is not limited to frequency measurement; an empirical model is also developed to estimate RC parasitic. It can correlate silicon measurement with simulation results as a practical way to identify the source of mismatch for improving interconnect mismatch.

## 3.2 Empirical Model

Compared with conventional ring oscillator approaches, the proposed test structure does not only predict the interconnect behavior through frequency measurement; the RC parasitic can be calculated through a set of ring oscillators and correlate with results with silicon measurements.

The empirical model is derived as follows:

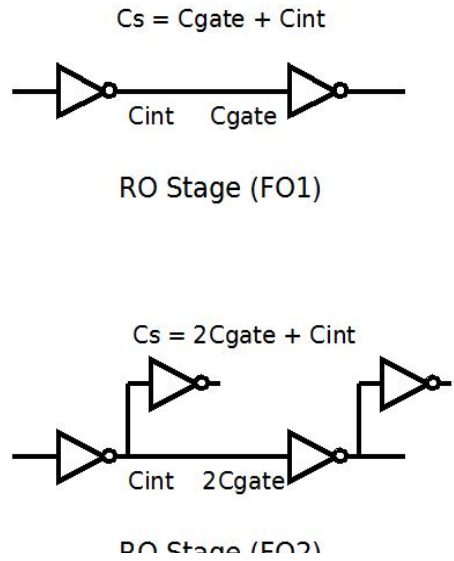

Figure 5 Reference and 2 Fanout RO Stage

Typically, the delay is expressed in term of supply voltage, average current and load capacitance as

$$T = C\frac{V}{I} \qquad \text{Eq. 11}$$

where it is also rewritten in term of PMOS/NMOS saturation current

$$T = CV\left(\frac{1}{I_{dp}} + \frac{1}{I_{dn}}\right) \qquad \text{Eq. 12}$$

where $I_{dp}$ is PMOS $I_{dsat}$ and $I_{dn}$ is NMOS $I_{dsat}$

Each stage switches twice during the complete cycle and the ring oscillator delay is calculated as follows:

$$f_d = \frac{1}{2nT_s} \qquad \text{Eq. 13}$$

where $f_d$ is the ring oscillator output frequency, n is the number of inverter stages and $T_s$ is the stage delay.

The ring oscillator delay $T_{osc}$ with output down counter scaling factor m is defined as

$$T_{osc} = mT_d \qquad \text{Eq. 14}$$
$$T_{osc} = 2nmT_s \qquad \text{Eq. 15}$$

The stage capacitance $C_s$ can be estimated using Eq. 11

$$C_s = \frac{T_s I_{eff}}{V_{dd}} \qquad \text{Eq. 16}$$

where $I_{eff}$ is effective current and is defined as the difference between the active current ($I_{dda}$) and leakage current ($I_{ddq}$) (i.e. $I_{eff} = I_{dda} - I_{ddq}$).

If $I_{ddq}$ is quite small, it is ignored during the calculation; then Eq. 16 is rewritten as

$$C_s = \frac{T_{osc} I_{eff}}{2mnV_{dd}} \qquad \text{Eq. 17}$$

From Eq. 17, the stage capacitance is calculated using in-phase crosstalk delay and current measurement rather than the shielded one because it eliminates coupling capacitance impacts.

The time delay is calculated with the switching resistance $R_{sw}$ and stage capacitance $C_S$ by

$$T_s = R_{sw} C_s \qquad \text{Eq. 18}$$

It can be simplified as

$$R_{sw} = \frac{V_{dd}}{2\, I_{eff}} \quad \text{Eq. 19}$$

Since the crosstalk introduces voltage noise on the victim, the overall delay and current measurement is changed; then, the quiet one measurement is chosen for switching resistance calculation.

The input gate capacitance $C_{gate}$ and output interconnect capacitance $C_{int}$ can be further estimated using two sets of ring oscillator with single fanout (FO1) and double fanout (FO2).

The stage capacitance $C_s$ can be divided into input $C_{in}$ and output one $C_{out}$ shown in Figure 5

$$\frac{T_{osc1}}{2mn} = R_{sw}(C_{int} + C_{gate}) \quad \text{Eq. 20}$$

$$\frac{T_{osc2}}{2mn} = R_{sw}(C_{int} + 2\, C_{gate}) \quad \text{Eq. 21}$$

To simplify Eq. 20 and Eq. 21 to obtain $C_{in}$ and $C_{out}$ as

$$C_{gate} = \frac{T_{osc2} - T_{osc1}}{2mnR_{sw}} \quad \text{Eq. 22}$$

$$C_{int} = \frac{2\, T_{osc1} - T_{osc2}}{2mnR_{sw}} \quad \text{Eq. 23}$$

With aggressor zero or step inputs, the victim output is evaluated through inverse Laplace Transform and expressed in Eq. 24 (in-phase), Eq. 25 (out-of-phase) and Eq. 26 (quiet mode) crosstalk operation:

$$V_i(t) = \left(1 - e^{-\frac{t}{RC}}\right) V_{dd} \quad \text{Eq. 24}$$

$$V_o(t) = \left(1 + \frac{2}{3}e^{-\frac{t}{RC}} - \frac{2}{3}e^{-\frac{t}{R(C+3C_c)}}\right)V_{dd} \qquad \text{Eq. 25}$$

$$V_q(t) = \left(1 - \frac{1}{3}e^{-\frac{t}{RC}} - \frac{2}{3}e^{-\frac{t}{R(C+3C_c)}}\right)V_{dd} \qquad \text{Eq. 26}$$

Eq. 24-26 are further simplified using the Taylor series expansion. Since RC constant is quite small, the second and higher coefficients are ignored for first-order approximation. Moreover, the coupling capacitance $C_c$ is much higher than the capacitance C (sum of input gate capacitance $C_{gate}$ and interconnect top/bottom capacitance: $C_{top}/C_{bottom}$). The term $R(C+3C_c)$ is rewritten as $3RC_c$. Moreover, the scaling factor ½ is taken into consideration to convert the lump model into a distributed one in order to match the simulation results. Finally, the capacitance C and $C_c$ are calculated as follows:

$$C = \frac{T_o T_q}{R(T_o + T_q)} \qquad \text{Eq. 27}$$

$$C_c = \frac{2T_o T_q}{3R(2T_o - T_q)} \qquad \text{Eq. 28}$$

where $T_o$ is the out-of-phase time delay and $T_q$ is the quiet one.

The empirical model can estimate the first order interconnect RC parasitic through simple measurements from our test structure. It is useful to monitor in-die and die-to-die RC parasitic variation and provide feedback to identify the source of variation during real chip production.

# Chapter 4. Experimental Results

## 4.1 Test Structure Characterization

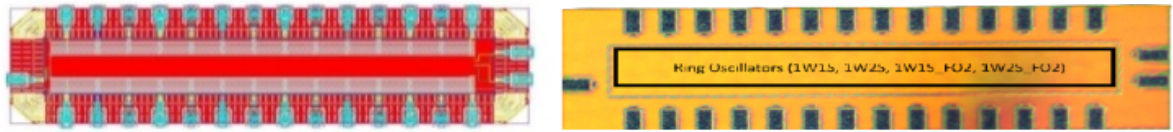

*Figure 6 Ring Oscillator Design*

In order to validate the test structures, four ring oscillators (RO) are implemented and taped out using TSMC 28nm HPM process as shown in Figure 6. They are divided into single width single spacing (1W1S) and single width double spacing (1W2S) ring oscillator with single & double fanout (FO1 & FO2) where single width and single spacing are referred to minimum line width and spacing. Every ring oscillator consists of control units and 100 inverter stages, and every inverter drives 50umlong Metal3 wire. The low Vt clock inverter is employed to minimize the insertion delay impacts and thelarge number of inverter stages are used to minimize the local device process variation. The 50um long Metal3 wire is chosen to fit the test chip area requirement with typical metal fill to achieve the density requirement.

For typical interconnect process, it is divided into three different metal configurations: the thin metal layer (Mx), the intermediate metal layer (My) and thick metal layer (Mz). The thin metal layer is often utilized for standard cell design and local routing (< 20um). The intermediate metal is targeted for global routing (> 20um), especially for clock tree design. The thick metal layer is used for Power Distribution Network (PDN) to minimize IR drop. Therefore, the intermediate metal layer (i.e. Metal3) is chosen for test chip implementation.

The current test chip structure has been modified for TSMC 7nmFF interconnect process evaluation, the actual wire width, spacing and length are revised to meet the transition time, duty cycle, crosstalk, jitter and electromigration requirements. The metal density variation is also taken into consideration due to high interlayer coupling as well as Chemical-Mechanical Planarization (CMP) effect. For TSMC 7nm FF process. The overall system performance is not only dominated by the device but alsointerconnect. Moreover, the test structures are used as figure of merit to compare the interconnect performance among different foundries. Based on preliminary simulation with internal enhanced interconnect model, the modified test structures have identified few TSMC 7nm FF interconnect issues and incorporated into critical path design.

Figure 7 shows that the output waveforms of various test structures: quiet, in-phase and out-of-phase one. The out-of-phase delay is longer, followed by quiet delay, then the in-phase one

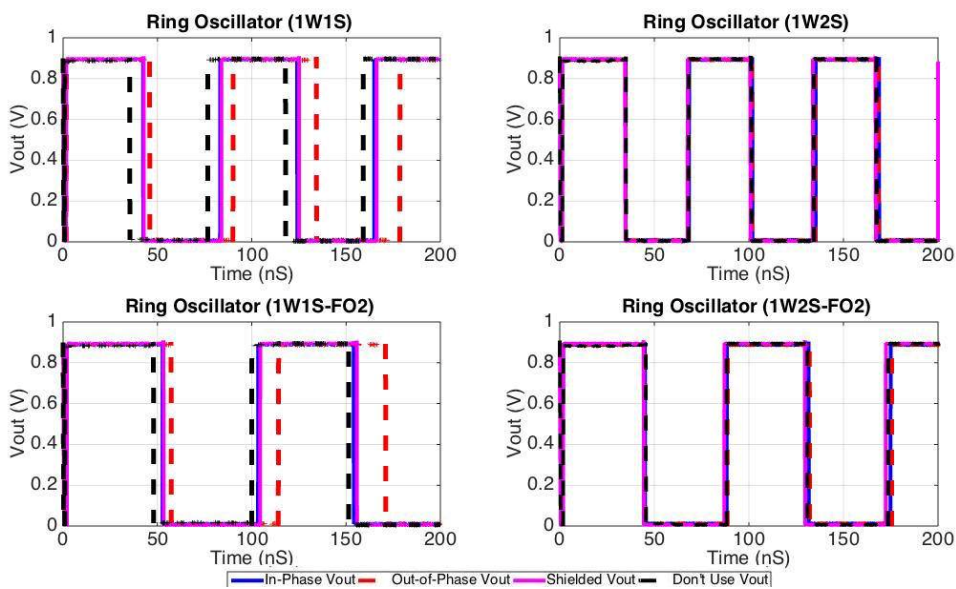

*Figure 7 Ring oscillator waveforms with different configuration*

The actual stage delay is highly related to the wire spacing. For single width single spacing configurations, the delay is improved with in-phase and out-of-phase crosstalk. With the double spacing, it reduces the coupling capacitance; the overall delay is similar among the three test structures. The measured RO time period ($T_{osc}$) and current ($I_{eff}$) results are shown in Table-1. The current ($I_{eff}$) is the difference between the normal current ($I_{dda}$) and the leakage one ($I_{ddq}$)

| Ring Oscillator | In-Phase Mode | | Out-of-Phase Mode | | Quiet Mode | |
|---|---|---|---|---|---|---|
| | Tosc (ns) | Ieff (uA) | Tosc (ns) | Ieff (uA) | Tosc (ns) | Ieff (uA) |
| 1W1S (FO1) | 81.66 | 891.50 | 88.39 | 990.63 | 82.31 | 503.47 |
| 1W1S (FO2) | 101.35 | 1388.00 | 113.22 | 1384.87 | 102.43 | 778.20 |
| 1W2S (FO1) | 65.99 | 1079.07 | 66.87 | 1093.13 | 66.22 | 532.07 |
| 1W2S (FO2) | 86.5 | 1518.57 | 87.00 | 1534.97 | 85.32 | 817.23 |

Table-1 RO Simulation Results

Through various RO structures, the RC parasitic can be calculated as shown in Table-2. We compare the data measured from our teststructure with that from the state-of-art test structure [31].Note that [31] can only measure quiet mode due to the lackof cross-talk aggressors. State-of-art test structure [31]estimates the total stage capacitance and switchingresistanceand cannot estimate $C_{int}$, $C_c$ and $C_{gate}$. Using newtest structures (i.e. in-phase and out-of-phase crosstalk),$C_{int}$, $C_c$, $C_{gate}$, $C_{total}$ and $R_{sw}$ can be calculated.

| Parameter | 1W1S | | | 1W2S | | |
|---|---|---|---|---|---|---|
| | this work | [19] | Spec | this work | [19] | Spec |
| Ctotal (fF) | 12.51 | 14.24 | 12.39 | 12.24 | 12.12 | 10.68 |
| Cgate (fF) | 3.02 | N/A | 2.54 | 3.82 | N/A | 2.54 |
| Cint (fF) | 9.50 | N/A | 9.85 | 8.42 | N/A | 8.14 |
| Cc(fF) | 6.82 | N/A | 7.91 | 6.81 | N/A | 5.51 |
| Rsw (Ω) | 504 | 497 | 450 | 417 | 423 | 276 |

Table-2. Model/Specification Comparisons

From the above tables, it is shown that our empirical model prediction is close to the design specification using various crosstalk modes. There are 1% (1W1S) and 15% (1W2S) Ctotal errors as well as 12% (1W1S) and 51% (1W2S) $R_{sw}$ errors between measurement and specification. The larger error of capacitance estimation in 1W2S is due to the fact that oscillation period in 1W2S is smaller than in 1W1S, and relative measurement error for oscillation period increases in 1W2S. Moreover, the large error may be related with double spacing OPC and CMP operation resulted in line width, spacing and thickness changes in actual silicon fabrication. With approach in [31], $C_{total}$ errors are around 14% and $R_{sw}$ errors are similar with proposed model. The total delay estimation errors of proposed test structure are 12% and 73% for 1W1S and 1W2S cases, while errors of [31] are 27% and 74% for 1W1S and 1W2S case. The results are further emphasized the proposed empirical model has a significant improvement toward the state-of-art test structure especially in small wire spacing case.

Due to limit sample available, the current test chip is targeted for parasitic validation only, however, the test structure can bridge gap between design and process for parasitic evaluation. It can befurther implemented in test vehicle or real chip (i.e. scribe line) to measure the interconnect process variation and identify potential process failure mechanism.

## 4.2 System Impact

To further validate the efficacy of the proposed teststructure in real designs, weapplyit as an interconnectmonitor to our in-house NPU taped out with 28 nm process. The NPUs implements AlexNet [32] forimage recognition. The chip area is 9mm x 9mm. It has 7hidden layers, 650,000 neurons and 60,000,000 weightparameters. The delay profile of interconnect can varysignificantly among chips located in different corners of thewafer. The die photo is shown in Figure 8. As can beseen from the figure, the proposed test structure can beeasily embedded by effectively taking advantage of thespare area between the two chips and accordingly areaoverhead is minimal.

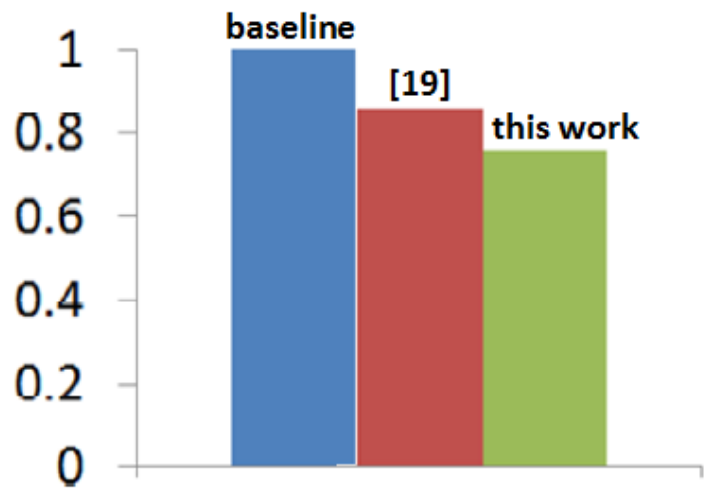

*Figure 8 Simulated total training time (normalized to baseline slow case)*

We measured the training time of NPU in various process corners with different interconnect delay, and the results are depicted in Figure 9 for two chips: the slowest one measured and one that can potentially run faster. Note that we have normalized the total training time with regards to the baseline worst case. Based on our measurement, the performance of NPU varies significantly with changes in interconnect and transistor delay, which can be as large as 30% between the best and the worst conditions. We estimated parasitic RC of chips in different corners by our on-chip monitor. We also used quiet-mode data to estimate the parasitic that would be reported by state-of-art test structure [31]. With conservative design methodology, we have to make the NPU work at the slowest clock frequency regardless of the actual interconnect delay, so the training time is always the same as the worst case baseline. With state-of-art test structure [31], the gate capacitance and interconnect delay can be tracked but with large error (~27%). Therefore, the clock rate estimated from state-of-art test structure [31] is still suboptimal, even though it can achieve better performance than baseline (~15% improvement). With our test structure, we can track transistor and interconnect delay variation accurately, and clock frequencycan be obtained. A performance improvement of 25% is obtained compared with baseline case.

# Chapter 5. Conclusions

Since interconnect RC parasitic plays an important role to predict the overall chip performance for nanotechnology, enhanced test structures are implemented to measure the routing propagation delay and crosstalk for various interconnect configurations. A first order empirical model is also developed to estimate RC parasitic. The test structures are quite simple and small. It can be easily inserted into real chip to monitor interconnect variation involume production. Recently, we have modified the proposed test structures to not only account for intra-layer (left/right routing) crosstalk but also for the interlayer (top/bottom routing) one. Additional device parameters (i.e. source/drain capacitance) areextracted via minor test structure update. The modified test structures and empirical models are useful to estimate both device and interconnect parameters, it also help us to identify the serious interconnect process issues for clock tree design using TSMC 7nm FF process.